\documentclass[conference]{IEEEtran}

\usepackage[T1]{fontenc}
\ifCLASSINFOpdf
  \usepackage[pdftex]{graphicx}
  \graphicspath{{figure/}}
  \DeclareGraphicsExtensions{.pdf,.jpeg,.png,.jpg}
\else
\fi
\usepackage{amsmath}
\providecommand{\keywords}[1]{\textbf{\textit{Keywords--- }}#1}
\usepackage[pdftex]{hyperref}
\begin{document}
%\IEEEoverridecommandlockouts
%\IEEEpubid{\begin{minipage}{\textwidth}\ \\[12pt]
% 2022 23rd International Conference on Electronic Packaging Technology (ICEPT)
%\end{minipage}}
%\title{Bare Demo of IEEEtran.cls\\ for IEEE Journals}
\title{\textit{Crystal structure informed mesoscale deformation model for HCP Cu$_6$Sn$_5$ intermetallic compound  }}
%\author{\IEEEauthorblockN{Anil~Kunwar}
%\IEEEauthorblockA{\textit{Faculty of Mechanical Engineering } \\
%\textit{Silesian University of Technology}\\
%Konarskiego 18A, 44-100 Gliwice, Poland \\
%Corresponding author, anil.kunwar@polsl.pl}
%\and
%\IEEEauthorblockN{Haoran~Ma}
%\IEEEauthorblockA{\textit{School of Microelectronics} \\
%\textit{Dalian University of Technology}\\
%116024 Dalian, China \\
%Corresponding author, mhr@dlut.edu.cn}
%\and
%\IEEEauthorblockN{Johan~Hektor}
%\IEEEauthorblockA{\textit{ Department of Materials Science and Applied Mathematics} \\
%\textit{Faculty of Technology and Society} \\
%\textit{Malmö University}\\
%21119 Malmö, Sweden \\
%johan.hektor@mau.se}
%}
\author{\IEEEauthorblockN{Anil~Kunwar$^{1,*}$, Haoran~Ma$^{2,**}$, Johan~Hektor$^{3}$} \\
\IEEEauthorblockA{$^1$Faculty of Mechanical Engineering, Silesian University of Technology, Konarskiego 18A, 44-100 Gliwice, Poland \\ $^2$ School of Microelectronics, Dalian University of Technology, 116024 Dalian, China \\ $^3$Department of Materials Science and Applied Mathematics, Faculty of Technology and Society, \\Malmö University,
21119 Malmö, Sweden \\ $^{*}$Email (Corresponding Author): anil.kunwar@polsl.pl ; $^{**}$Email (Corresponding Author): mhr@dlut.edu.cn}}

\maketitle

\begin{abstract}
%The abstract goes here.
%As the electronic packaging and fabrication industries for solar PV cells assembly or panels are  making endeavours in using Pb-free solder joints, the reliability of these joints can be taken as an important research topic. Among several aspects, the control of void formation, growth and evolution in lead-free Sn-based solders  can be achieved if the underlying mechanism for such phenomena can be modeled. This study employs a quantitative polynomial free energy based phase field method to model the motion and evolution of void in Sn material under thermal gradient. The effects of imposed temperature gradient to the overall migration rate and profile of the void has been assessed in the finite element model.
In the electronic packaging and energy storage sectors, the study of Cu$_6$Sn$_5$ intermetallic compound (IMC) is getting more attention. At temperatures above 186 $^\circ$C, this IMC exists in a hexagonal closed packed (HCP) crystalline structure. Crystal plasticity finite element simulations are performed on Cu$_6$Sn$_5$ IMC by taking the information about its lattice parameters and direction dependent elastic properties. Three types of models corresponding to deformations in basal, prismatic and pyramidal modes are developed. With the same type of loading in the elastic regime and boundary conditions, the results of the computations reveal the differences in displacement magnitudes among the  three model types.
\end{abstract}

%\begin{IEEEkeywords}
%IEEE, IEEEtran, journal, \LaTeX, paper, template.
%\end{IEEEkeywords}

% \fi
\IEEEpeerreviewmaketitle

\section{Introduction} \label{intro_1}
The use of Pb-free solders joints is getting promotion and significant research interest because of the immense benefit to environment and human health \cite{Dong2022a,Kunwar2020x}.  For Sn-based solder joints using Cu substrates, the formation of joint is guaranteed by the growth of Cu$_6$Sn$_5$ intermetallic compound (IMC) at the interface. For miniaturized (Pb-free) Sn-based solder joints, the volume fraction of the Cu$_6$Sn$_5$ IMC  in the overall composite joint can be proportionately significant. For an example, in context of 2.5D IC packaging, the diameters of solder balls can be considered in the values well below 10 $\mu$m, and so during the reflow joining procedure with Cu substrate, the major portion of the liquid solder is consumed for the interfacial reaction and formation of the IMC.  This  suggests that the knowledge of the structure and  properties of  Cu$_6$Sn$_5$ IMC  can eventually help in the design of strong and reliable solder joints.  Besides the electronic packaging sector, the Cu$_6$Sn$_5$ IMC is known as a promising candidate materials for the electrode of the Lithium-ion battery \cite{Tan2022a}.
\begin{figure}
\includegraphics[scale=0.09]{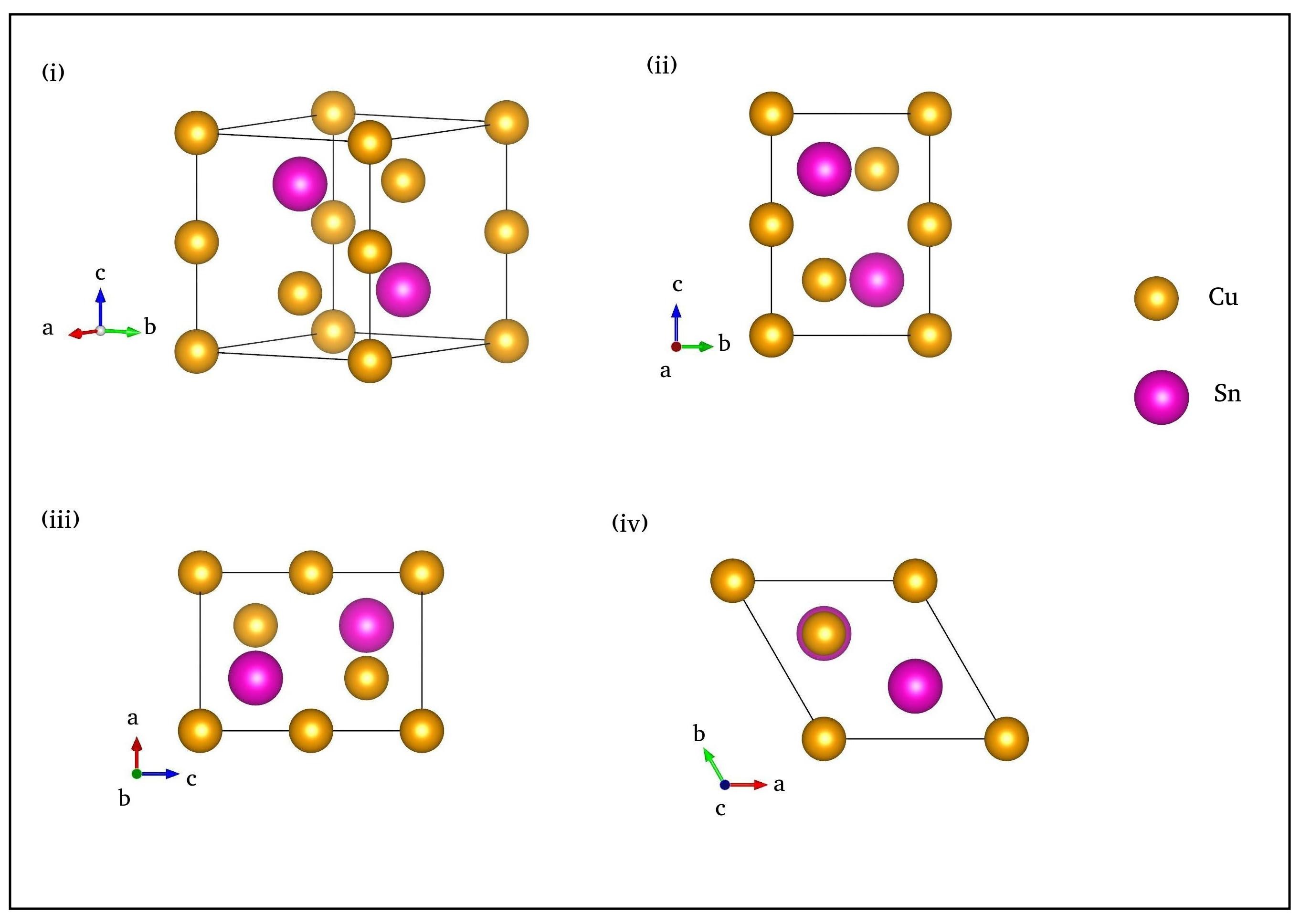}
\caption{Perspective view of a unit cell of $\eta$-Cu$_6$Sn$_5$ compound is presented in (i). The images in (ii), (iii) and (iv) respectively show the view along the planes normal to respectively  <a>,  <b> and  <c> directions.}
\label{fig-1}
\end{figure}  
%\paragraph*{}
\par Cu$_6$Sn$_5$ IMC  can exist in two forms of crystalline structures \cite{Mu2016}. At temperatures below 186 $^\circ$C \cite{Tan2022a}, it has a monoclinic crystalline structure, and is referred to as $\eta'$-Cu$_6$Sn$_5$. However, for temperatures higher than  186 $^\circ$C, it undergoes a structural transformation to have a hexagonal closed packed structure \cite{Liang2022a,Hektor2016c}. The IMC with hexagonal closed packed (HCP) crystalline structure is also termed as  $\eta$-Cu$_6$Sn$_5$. During reflow soldering procedure, the Sn-based solder has to undergo melting and as the melting point of pure Sn is 523.15 K, $\eta$-Cu$_6$Sn$_5$ is formed during the interfacial reaction between the solder and substrate. In this study, $\eta$-Cu$_6$Sn$_5$ is considered as the material for computational analysis. The currently widespread applications such as high-powered electronic equipment and third generation semiconducctor power devices have  high service temperature ( T > 523.15 K) requirements ~\cite{Ding2021}, and this makes the study of   $\eta$-Cu$_6$Sn$_5$ phase more relevant.
%The different types of solder voids are outlined in the work of Liang et al \cite{Liang2016}. Presence of voids in Sn-based solders creates  problems such as reduction of effective area of the joints, acts as a region of stress concentration for mechanical loads, alters the thermal and electrical properties of the materials and makes the system vulnerable to mechanical failure. In other words, the voids have significant influence on the integrity and performance of solder interconnects \cite{Liang2015}. In the presence of electrical and thermal field gradients, the numerical analysis of motion and shape evolution of void is deemed very important as the experimental prediction of such phenomena is difficult owing to the very small size of the solder joint. In this study, a phase field model is developed for a void evolution in Sn with thermomigration as the driving force. Owing to the capability of quatitative two phase polynomial free energy over thermodynamic free energy in simplying the numerical solutions \cite{Tonks2013},a quantitative polynomial free energy based phase field model has been utilized  in this work to describe the free energy of Sn material containing void.
\section{Computational study of the IMC} \label{mathematical_model2} 
\subsection{Crystal structure} \label{hcp_geometry}
\begin{figure}
\includegraphics[scale=1.0]{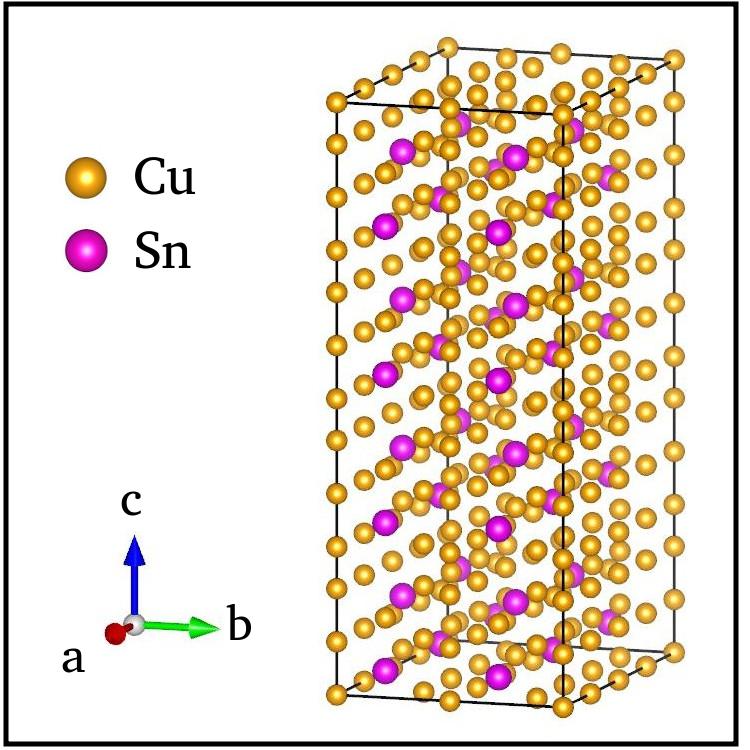} %0.2025
\caption{ The 2 $\times$ 2 $\times$ 4 supercell of the unit cell presented in Fig. 1 (i).}
\label{fig-2}
\end{figure}

The location of the the Cu atoms and Sn atoms as viewed on the unit cell of HCP Cu$_6$Sn$_5$ compound is presented in Fig. ~\ref{fig-1} and that for the 2 $\times$ 2 $\times$ 4 supercell is provided in Fig. ~\ref{fig-2}. The visualization is done using the Visualisation for Electronic Structural Analysis (VESTA) software \cite{Momma2011}. The lattice parameters for the HCP crystal are a = b = 4.190 {\AA}, c=5.086 {\AA}, $\alpha$ = $\beta$ = 90 $^\circ$, and $\gamma$ = 120 $^\circ$ \cite{Mu2016}.  The lengths a, b and c can are the dimensions of the unit cell corresponding to <a>,<b> and <c> directions respectively. In Fig. ~\ref{fig-1}(i),the height of the unit cell is 5.086 {\AA}.The views provided in Fig. ~\ref{fig-1}(ii)-(iv) illustrate the angles $\alpha$, $\beta$  and $\gamma$.
%Subsection text here.
%The two dimensional (2D) geometry and mesh representing the domain of the numerical simulation is presented in Fig. ~\ref{fig-1}.  The rectangular pure tin, consisting of an initially circular void of several nanometers, is modelled as a computational domain  of length(X$_{max}$) = 1.00 $\mu$m and breadth(Y$_{max}$) = 0.5 $\mu$m . The element type of two dimensional mesh element was selected as quadrilateral (QUAD). The size of an individual mesh element ($ \Delta~x \times \Delta~y$) is set as 3.33 nm $\times$ 1.67 nm.  For phase field modeling, the initial centre point position of a circlular void was specified at the left side of the computational domain. The values of phase field variable c were respectively chosen as 1 and 0 for the void and metal Sn. The radius (R) of the circle is initially assigned as  150 nm. For the sake of numerical accuracy, the following conditions \cite{Durga2015,Moelans2008,Moelans2008a} are maintained regarding the width or thickness (l$_i$) of the diffuse interface (0 < c <1) :

%%%%%%%%%%%%%%%%%%%%%%%%%%%%%%%%%%%%%%%%%%%%%%%%%%%%%%%%%%%%%%%%%%%%%%%%%%%%%
%%%%%%%%%%%%%%%%%%%%%%%%%%%%%%%%%%%%%%%%%%%%%%%%%%%%%%%%%%%%%%%%%%%%%%%%%%%%%
\subsection{Elastic properties of $\eta$-Cu$_6$Sn$_5$ } \label{elasticity_tensor}
\begin{figure}[b]	%[h]
 \centering
\includegraphics[scale=0.350]{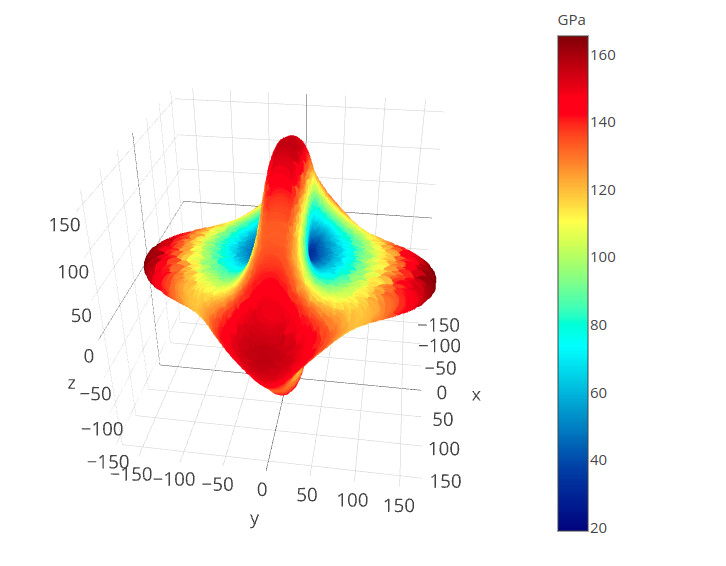}
\caption{Visualization of directional dependence of Young's modulus of Cu$_6$Sn$_5$ intermetallic compound.}
\label{fig-3}
\end{figure}

\begin{figure}[h]
 \centering
\includegraphics[scale=0.32525205]{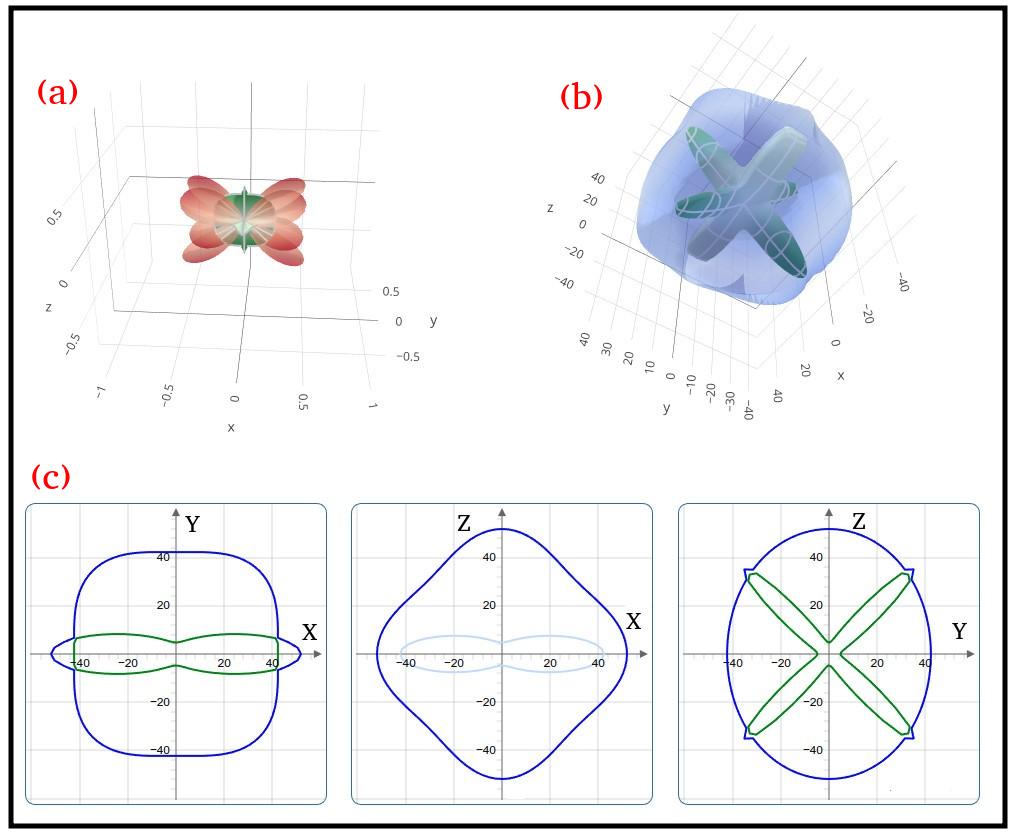}
\caption{Assuming the IMC as an orthotropic material, the directional dependence of poisson's ratio is illustrated in (a) whereas the spatial variation of shear modulus is presented in (b)-(c). The image in (c) corresponds to the 2D visualization of the 3D image provided in (b). Th unit corresponding to the numerical values represented in (b) and (c) is GPa.}
\label{fig-4}
\end{figure}

The $\eta$-Cu$_6$Sn$_5$, with HCP structure, is characterized by anisotropy in elastic properties. The elasticity (stiffness) tensor (C$_{ijkl}$) represents the elastic behavior of a crystalline matrix. For simplifying  the models,  different materials symmetry options are utilized. In this work, orthotropic symmetry is assumed for the IMC crystal. The values for the 9 components of the elasticity tensor of the IMC crystal are obtained from the work of Lee et al \cite{Lee2006a}. From the stiffness tensor, it is possible to compute and visualize direction dependent Young's modulii, Poisson's ratios and shear modulii. The images corresponding to the visualization of Young's modulii, Poisson's ratios and shear modulii are presented in Figs. ~\ref{fig-3} and ~\ref{fig-4}.

\par The visualization of Young's modulii (Fig. ~\ref{fig-3}) is done with the help of Multi-phase Elastic Aggregates (MELASA) software \cite{Friak2020}. From the figure, it can be understood that the Young's modulus of  Cu$_6$Sn$_5$ IMC crystal is direction dependent, and its magnitude fluctuates in the range 18.15-131.75 GPa. The visualization of Poisson's ratios and shear modulii (Fig. ~\ref{fig-4}) is performed using ELATE software \cite{Gaillac2016a}.  The 3D image for variation of Poisson's ratio along the different directions is shown in Fig. ~\ref{fig-4}(a). The magnitudes of the Poisson's ratio are in the range from 0.03 to 0.82.  The image showing the variation of shear modulus along different direction is presented as 3D figure in Fig. ~\ref{fig-4}(b)  and the corresponding projections on 2D planes in Fig. ~\ref{fig-4}(c). The minimum value of shear modulus is 4.8 GPa and the maximum magnitude is 51.4 GPa. 

\section{Crystal plasticity finite element simulation} \label{results_discussions1}
%Subsection text here. about
In context of finite strain inelastic mechanics of crystal plasticity, the deformation gradient (\textbf{F}) in IMC is assumed to be composed of elastic deformation gradient (\textbf{F$^e$})  and plastic  deformation gradient (\textbf{F$^p$}), as following:
\begin{equation} \label{constitutive_relationship}
\textbf{F} = \textbf{F$^e$}\textbf{F$^p$}
\end{equation}
where, det(\textbf{F$^e$}) > 0 and det(\textbf{F$^p$}) = 1.
With Cauchy stress denoted as $\sigma$, the following equation can be utilized to represent the Second Piola-Kirchhoff Stress:
\begin{equation}
\textbf{S} = det(\textbf{F}) \textbf{F$^{-1}$} \sigma \textbf{(F$^{-1}$)$^{T}$}
\end{equation}
\par The HCP crystal of  $\eta$-Cu$_6$Sn$_5$ consists of three slip modes - (i) basal (slip system \{0001\} <11$\overline{2}$0>), (ii) prismatic (slip system \{10$\overline{1}$0\}
<11$\overline{2}$0> ) ; and <a> pyramidal (slip system \{10$\overline{1}$1\}
<11$\overline{2}$0>) , and these modes provide deformation only  to <a> direction \cite{choudhury2015a}. In addition to this, the  deformation can can occur via combined <c+a> pyramidal slips (slip systems \{10$\overline{1}$1\}
<11$\overline{2}$3> and \{10$\overline{2}$2\} <11$\overline{2}$3> ) and deformation twinning.
\par In this work, the deformation behavior along basal, prismatic and pyramidal modes is inspected for an IMC block. The finite element method based numerical simulations were performed in Multiphysics Object Oriented Simulation Environment (MOOSE) software \cite{Permann2020a,Pitts2022}. A cube of dimension (1 mm $\times$ 1 mm $\times$ 1 mm) is chosen as the geometry and the mesh is constructed on this geometry. A total number of 216 HEX8 mesh elements were assigned in the cube mesh. Appropriate boundaries were assigned to establish the three modes. A  strain rate of 5.0E-4 s$^{-1}$ was applied to one face of the cube  for all of the three simulations. The elasticity tensor was supplied in the Materials block, and symmetric9 fill method is utilized to consider orthotropic symmetry of the IMC material. The information of slip systems and  lattice constants (a, b and c) of the HCP crystal for Cu$_6$Sn$_5$ IMC also was provided in the Materials block. For the current simulation, the grain size of 15 $\mu$m is considered for the IMC grains in the cube. The temperature of the computational domain is considered to be equal to 470.0 K.
\par The yield strength of Cu$_6$Sn$_5$ IMC is 2009 MPa \cite{Deng2004}. Thus it should be noted that the loading described above will not be sufficient to initiate the plastic slip based deformation in the HCP crystal. Although this work will utilize the methodology of crystal plasticity finite element simulations, the following results and discussions sections will be confined within the elastic regime. That is, the stress discussed in this study will be so small that it will not induce any significant plastic deformations. The choice of elastic regime is favorable from the viewpoint of numerical convergence and computational efficiency.

\section{Results and discussions} \label{results_discussions2}
\begin{figure}[h]
\centering
\includegraphics[scale=0.750]{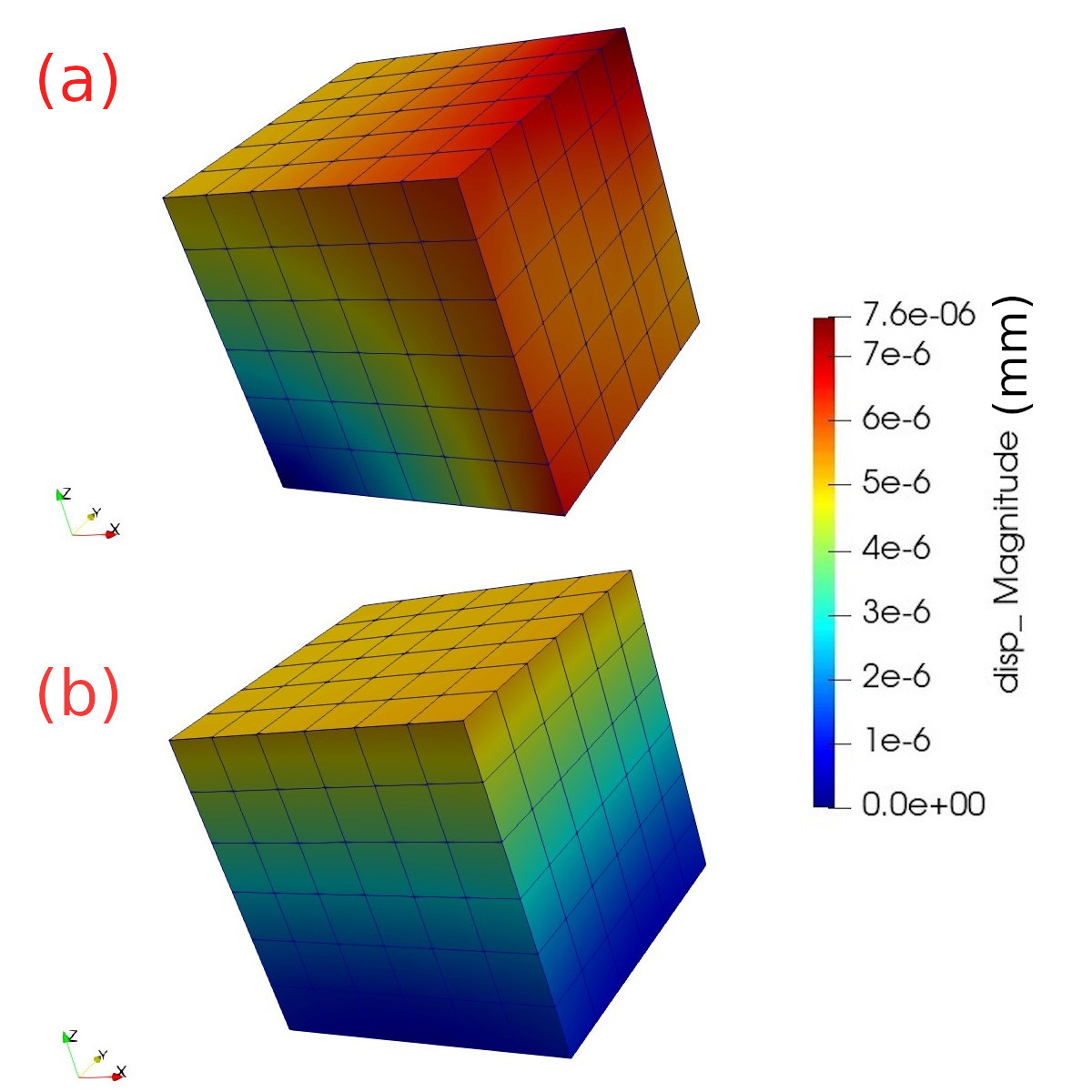}
\caption{Computed displacement at t=0.01 s for deformation slip along <a> direction for the (a) basal and (b) prismatic modes.}
\label{fig-5}
\end{figure}

\begin{figure}[h]
\centering
\includegraphics[scale=0.750]{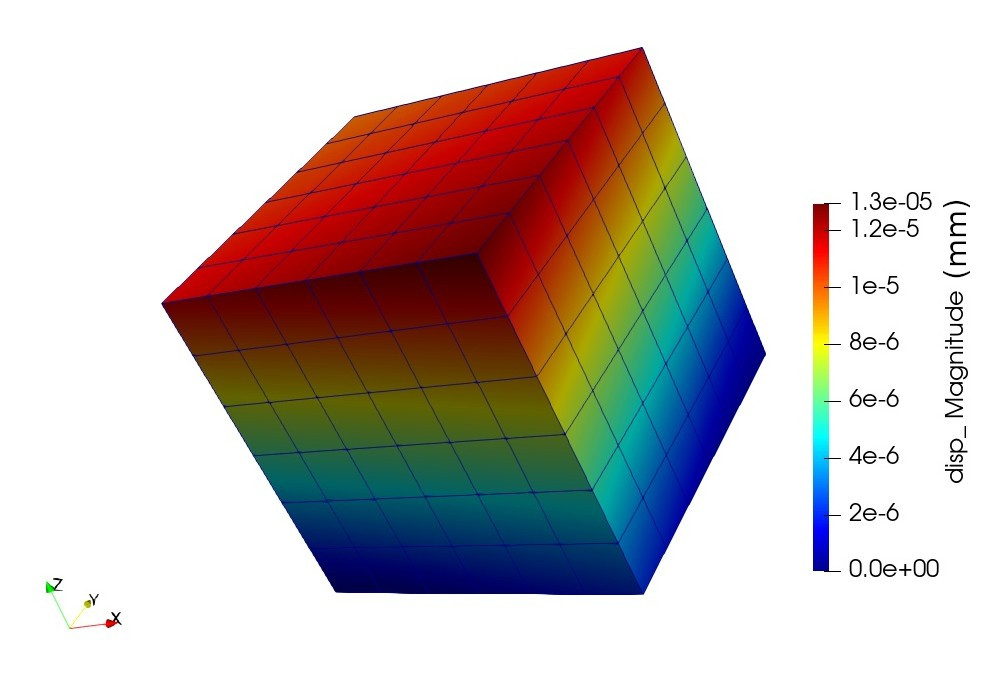}
\caption{Simulated displacement at t=0.01 s for deformation  along <c+a> direction for the pyramidal deformation mode.}
\label{fig-6}
\end{figure}

The results for displacements at t=0.01 s for simulations corresponding to basal and prismatic modes of deformation are presented in Fig. ~\ref{fig-5}(a)-(b). It has to be noted that for a given strain rate, the deformation  profiles for the two modes is different. The entire frontal part of the IMC cube for basal mode undergoes displacement with magnitudes larger than 6.0e-6 mm. Some points in the cube reach a displacement as high as 7.6e-6 mm. However, in case of prismatic mode, the maximum displacement reached at the top face of the cube is only 5.24E-6 mm. Moreover, instead of entire frontal region, the deformation is more prevalent in the top face of the cube.  The displacement magnitude  at t=0.01 s in context of deformation corresponding  to <c+a> pyramidal mode, is presented in Fig. ~\ref{fig-6}. As anticipated, for the same magnitude of applied strain rate, the cube is characterized by larger deformation at the top frontal region (as compared to two other modes). The maximum displacement magnitude at the top regimes is around 1.3E-05 mm . 
\par
\begin{figure}[b]	%[h]
 \centering
\includegraphics[scale=0.22350]{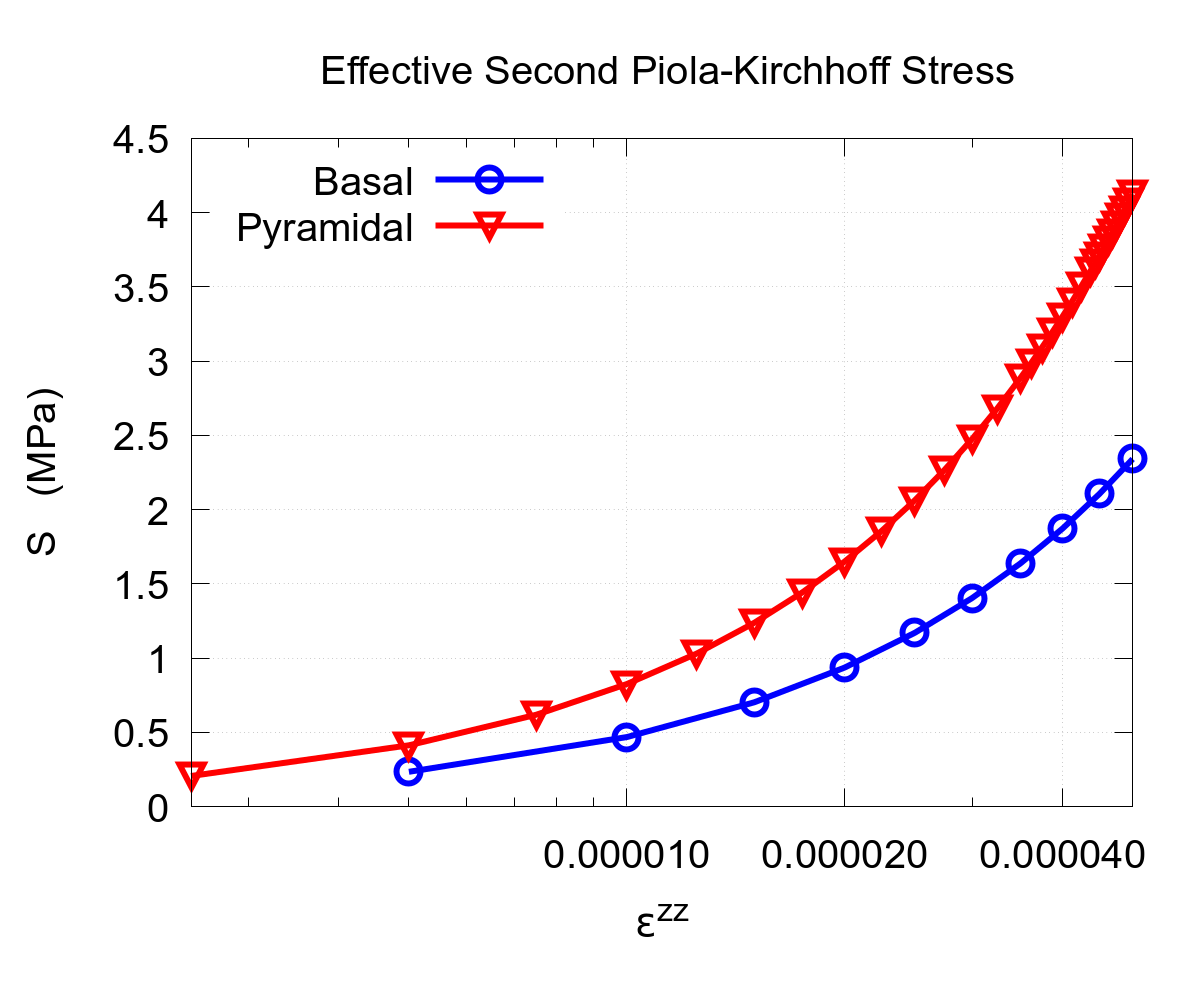}
\caption{Effective Second Piola-Kirchoff Stress versus strain along Z-direction for basal and pyramidal deformation modes.}
\label{fig-7}
\end{figure}

 For the loading in the [001] direction, it is important to compare the magnitudes of effective Second Piola-Kirchoff stress between basal and pyramidal modes, and such comparison is provided in Fig. ~\ref{fig-7}. As shown in the figure, when the strain along this direction is 5.0E-5, the average magnitude of stress for Basal mode is 2.5 MPa whereas the effective stress for Pyramidal mode is above 4 MPa.
These facts reveal that the information obtained from crystal structure when used together with mesoscale simulations, can provide  vast range of data about the materials behavior under external loading.
\par As illustrated by the magnitudes of stress above,it can be said that the current models are confined within the elastic regime. In context of applications of HCP Cu$_6$Sn$_5$  IMC undergoing elasto-plastic loadings; these existing models have the capability to incorporate loadings that are large enough to induce stresses beyond the yield strength of the material. In future, the numerical simulations will be performed in the plastic regime. 

\section{Conclusions} \label{conclusion_inference3}
%The conclusion goes here. 
The  following conclusions  have been derived from this study :
\begin{enumerate}
\item At temperature above 186 $^\circ$C, Cu$_6$Sn$_5$ IMC has a HCP structure. In this study, this structural information is utilized in the finite element method based deformation model at mesoscale.

\item The lattice constants and the elasticity  tensors of Cu$_6$Sn$_5$ IMC have been supplied to the mesoscale model as input materials parameters.  

\item Three different modes of deformation - basal, prismatic and pyramidal, are considered in the mesoscale model. The displacement magnitudes in the three modes are studied for a given strain rate of 5.0E-04 s$^{-1}$ in the Z-axis direction. It is found that the maximum displacement of the computational domain (Cu$_6$Sn$_5$ IMC cube) under basal mode is larger as compared to that of prismatic mode. While all regions of the cube undergo a displacement in prismatic mode for the given loading condition, it is possible for some local regions in basal mode to not undergo deformation. In pyramidal deformation mode, all locations of the cube undergo displacement; and the magnitudes are larger compared to the other two modes.
%The velocity of void's center is drifted a maximum x-displacement for the void of initial radius of 100 nm and thermal gradient of 197 k/$\mu$m for the simulation time duration of 25 $\mu$s. This confirms  that the drift velocity pore's center is directly proportional to the magnitude of thermal gradient whereas is inversely proportional to the initial size of the void. 

\item For a given strain magnitude, the average value of Second Piola-Kirchoff stress in pyramidal deformation mode is larger as compared to the basal mode. This facts illustrate the importance of using atomistic informations in mesoscale simulations.
%\item The accuracy in tracking the surface evolution of the void under temperature gradient is enhanced by using a greater $\frac{R}{l_{i}}$ ratio in phase field model.

\end{enumerate}

\section*{Acknowledgment}
This work was supported by the National Science Centre, Poland (Grant Number: 2021/42/E/ST5/00339)  and the National Natural Science Foundation of China (Grant Number:  52101035).

\ifCLASSOPTIONcaptionsoff
  \newpage
\fi



%\bibliographystyle{IEEEtran}
%\begin{thebibliography}{1}


%\end{thebibliography}
\bibliographystyle{unsrt}
%\bibliography{icept_2016}
\bibliography{elasticity_reference}


%\end{IEEEbiography}


%\newpage



%\vfill


\end{document}